\def\year{2021}\relax
\documentclass[letterpaper]{article} 
\usepackage{aaai21}  
\usepackage{times}  
\usepackage{helvet} 
\usepackage{courier}  
\usepackage[hyphens]{url}  
\usepackage{graphicx} 
\usepackage{csquotes}
\usepackage{amssymb}
\usepackage{rotating}
\usepackage{amsmath}
\usepackage{natbib}  
\usepackage{caption} 
\usepackage{array}
\usepackage{flushend}
\usepackage{setspace} 
\usepackage{makecell}
\usepackage[flushleft]{threeparttable}
\usepackage{booktabs,caption}
\usepackage[toc,page]{appendix}
\usepackage{subcaption}
\usepackage{rotating}
\usepackage{multirow}
\usepackage[T1]{fontenc}

\usepackage[usenames,dvipsnames]{color}

\title{Doctors vs. Nurses: Understanding the Great Divide in Vaccine Hesitancy among Healthcare Workers}
\author {
    Sajid Hussain Rafi Ahamed\textsuperscript{1}, Shahid Shakil\textsuperscript{1}, Hanjia Lyu\textsuperscript{1}, Xinping Zhang\textsuperscript{2}, Jiebo Luo\textsuperscript{1}\\
}
\affiliations {
    \textsuperscript{1} University of Rochester\\
    \textsuperscript{2} University of Rochester Medical Center\\
    \{srafiaha, sshakil2, hlyu5\}@ur.rochester.edu, xinping\_zhang@urmc.rochester.edu, jluo@cs.rochester.edu\\
}

\begin{document}

\maketitle

\begin{abstract}
Healthcare workers such as doctors and nurses are expected to be trustworthy and creditable sources of vaccine-related information. Their opinions toward the COVID-19 vaccines may influence the vaccine uptake among the general population. However, vaccine hesitancy is still an important issue even among the healthcare workers. Therefore, it is critical to understand their opinions to help reduce the level of vaccine hesitancy. There have been studies examining healthcare workers' viewpoints on COVID-19 vaccines using questionnaires. Reportedly, a considerably higher proportion of vaccine hesitancy is observed among nurses, compared to doctors. We intend to verify and study this phenomenon at a much larger scale and in fine grain using social media data, which has been effectively and efficiently leveraged by researchers to address real-world issues during the COVID-19 pandemic.  More specifically, we use a keyword search to identify healthcare workers and further classify them into doctors and nurses from the profile descriptions of the corresponding Twitter users. Moreover, we apply a transformer-based language model to remove irrelevant tweets. Sentiment analysis and topic modeling are employed to analyze and compare the sentiment and thematic differences in the tweets posted by doctors and nurses. We find that doctors are overall more positive toward the COVID-19 vaccines. The focuses of doctors and nurses when they discuss vaccines in a negative way are in general \textit{different}. Doctors are more concerned with the effectiveness of the vaccines over newer variants while nurses pay more attention to the potential side effects on children. Therefore, we suggest that more customized strategies should be deployed when communicating with different groups of healthcare workers.
\end{abstract}

\section{Introduction}

COVID-19, a novel coronavirus, was identified in December 2019 as a potential cause of respiratory infections such as cough, flu, and pneumonia. The WHO (World Health Organization) Emergency Committee declared a global pandemic on March 11, 2020, due to the rapid spread of the virus world wide~\cite{cucinotta2020declares}. As of August 29, 2022, there have been 602,479,782 infections and 6,480,637 deaths in total around the globe.\footnote{\url{https://covid19.who.int/}}

Vaccination has considerably decreased the burden of numerous infectious illnesses throughout history. Therefore, developing SARS-CoV-2 vaccines and making them available worldwide is a top goal for stopping the pandemic~\cite{lyu2022social}. \citet{khubchandani2021covid} show that 68\% of people are willing to take the vaccine and 22\% are disinclined toward the vaccine. Vaccine hesitancy indicates that vaccine availability does not guarantee adequate vaccination coverage. Previous research has shown that vaccine compliance is varied~\cite{lyu2022social}. Socioeconomically disadvantaged groups are more likely to hold polarized opinions on COVID-19 vaccines, either pro-vaccine or anti-vaccine~\cite{lyu2022social}. Health communications on social platforms about the advantage and safety of getting vaccinated does not necessarily increase vaccine uptake due to the negative impact of online social endorsement~\cite{lyu2022misinformation}. There are both challenges and opportunities in the process of designing effective communication and immunization programs~\cite{wu2021characterizing}. Effective vaccinations against COVID-19 would need massive public education campaigns about vaccine safety and efficacy~\cite{lucia2021covid}. Medical professionals are among the group of front-line healthcare providers who are likely to be exposed to COVID-19 patients. Trust in healthcare workers is associated with increased vaccination intention~\cite{de2020mapping}. Their opinions will be entrusted with providing vaccine recommendations and counseling vaccine-hesitant patients.\footnote{\url{https://www.usnews.com/news/national-news/articles/2019-01-16/who-names-vaccine-hesitancy-as-top-world-threat-in-2019}} Hence, research on vaccine hesitancy among healthcare workers is warranted. Multiple questionnaire-based studies have been conducted to examine the perceptions of the COVID-19 vaccines and willingness in vaccination among healthcare workers~\cite{adejumo2021perceptions, iliyasu2022should, grumbach2021association,kara2021covid, amuzie2021covid, nasir2021perception}. Interestingly, although doctors and nurses are both healthcare workers, there is a considerable difference in the level of vaccine hesitancy regarding COVID-19. A higher proportion of nurses are COVID-vaccine hesitant compared with doctors~\cite{browne2021coronavirus}. 

In the era of the COVID-19 pandemic, social media data has been leveraged by researchers for various purposes~\cite{tsao2021social} such as analyzing public attitudes~\cite{abd2020top, lyu2022social}, identifying infodemics~\cite{gallotti2020assessing, lyu2022misinformation}, assessing mental health~\cite{gao2020mental, zhang2021monitoring}, and so on. To our best knowledge, there have not been any social media studies on healthcare workers' opinions about COVID-19 vaccines. To fill this gap, our study focuses on using social media data to understand medical practitioners' opinions toward COVID-19 vaccines, the reasons for their hesitancy if any, and how this may translate to the general public's opinion, as these factors may influence the general public's vaccine decisions.  

Specifically, this study uses the keyword search in user's profile description to identify whether or not a Twitter user is a healthcare worker. Based on the keywords, the detected medical professionals are further divided into doctors, nurses, and others. We focus on comparing the sentiments and themes of the tweets posted by doctors and nurses. We find that doctors are more concerned with the ineffectiveness of the vaccines on the new variants while nurses pay more attention to the risk of vaccine side effects on children.  

\section{Related Work}
Vaccine hesitancy, defined as a reluctance to vaccinate despite the availability of vaccinations, is one of WHO's top 10 global health hazards for 2019.\footnote{\url{https://www.who.int/news-room/spotlight/ten-threats-to-global-health-in-2019}} Since then, several studies have analyzed the reasons for the public hesitancy toward various vaccinations.

\citet{khubchandani2021covid} launched a survey to assess the COVID-19 vaccine hesitancy in the community-based sample American population. The survey included 1,878 participants, most of whom were {\tt women}, {\tt Whites}, {\tt non-Hispanic}, {\tt married}, {\tt working full-time}, and had a {\tt bachelor's degree or above}. In the study population, the likelihood of receiving a COVID-19 vaccine was: very likely (52\%), somewhat likely (27\%), not likely (15\%), and not (7\%), with individuals with lower education, income, or perceived risk of infection being more likely to report that they were not likely to receive the vaccine. \citet{lyu2022social} proposed a human-guided machine learning framework to evaluate over six million tweets to categorize public opinion on COVID-19 vaccination and found that political affiliation and COVID-related experience are correlated with vaccination intention. 

There are a few studies on understanding opinions toward COVID-19 vaccination among particular groups of population. For instance, \citet{lucia2021covid} found that students who were willing to receive the vaccination right away were more inclined to believe public health officials, had fewer concerns about adverse effects, and supported vaccine requirements. They highlighted the need for an educational curriculum to improve students' awareness of COVID-19 vaccines and educate them on how to give vaccination advice. In terms of the vaccine hesitancy among medical professionals, \citet{adejumo2021perceptions} found that educational interventions to improve healthcare workers' perceptions and attitudes toward the COVID-19 vaccine are also necessary. There are several common reasons for not getting vaccinated. Healthcare workers may have doubt on the efficacy of the vaccine, distrust of its content, and fear of side effects~\cite{kara2021covid}. Building COVID-19 vaccine confidence may increase vaccine acceptance among them~\cite{iliyasu2022should}. Our study intends to investigate this question at a larger scale and in fine grain with social media data.

\section{Material and Method}
In this section, we describe how we collect Twitter data, identify medical professionals, and remove irrelevant data that do no disclose users' opinions with respect to COVID-vaccine uptake. We also discuss how we estimate and analyze the sentiments and themes in the online discussion of doctors and nurses. 

\subsection{Data Collection}
Following \citet{lyu2022social}, we use the Twitter API - Tweepy, to collect the publicly available tweets that contain particular keywords from September to November, 2021. The same keyword list of \citet{lyu2022social} is used which is composed of “vaccine”, “COVID-19 vaccine”, “COVID vaccine”, “COVID19 vaccine”, “vaccinated”, “immunization”, “covidvaccine”, “\#vaccine”, “covid19 vaccine”, “vacinne,” “vacine,” “antivax,” and “anti-vax.” We remove the tweets that contain “vaccine” but are not related to COVID-19. For example, these tweets may discuss flu shots or polio vaccines.

\begin{figure*}[htbp]
    \centering
    \includegraphics[width=0.85\linewidth]{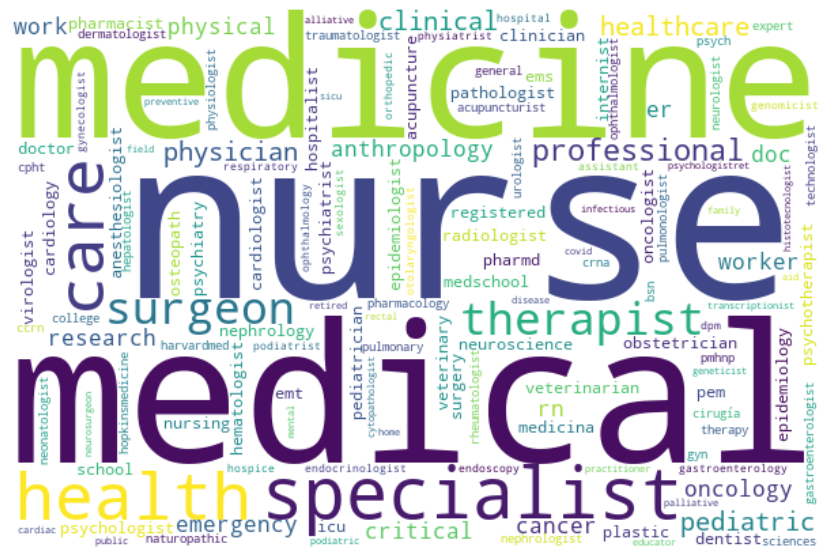}
    \caption{The keywords used for medical professional inference.}
    \label{fig:keywords_medical}
\end{figure*}

\subsection{Medical Professional Inference}
Twitter users can disclose their professions in profile descriptions. For example, a Twitter user may write 

{\tt "Wife Mother Retired Nursing"}

We use the keyword search to identify whether or not a user is a medical professional. First, we manually read the profile descriptions of 9,325 unique users and construct the keyword list related to the medical professions. We find 148 prominent keywords (Figure~\ref{fig:keywords_medical}) that identify the user's profession. If the profile description does not contain any of the keywords, it is labeled as non-medical professional. Some example keywords are “doctor”, “nurse”, “icu”, “rn”, “pathologist”, “bsn”, “anthropology”, “epidemiology”, “medicine”, “emt”, “virologist”, “nephrologist”, “pharmacist”, “cardiology”, “surgeon”, “pharmd”, “psychiatry”, “therapist”, etc. We apply the keyword search to the remaining non-labeled data. We test our keyword search method on the labeled dataset. Table~\ref{tab:keyword} shows the classification performance. In the sample of 152,213 unique users, 3,904 medical professionals are observed, accounting for about 3\% of the sample size. This result is very much \textit{in line with} the Census Bureau's 2019 American Community Survey,\footnote{\url{https://www.census.gov/programs-surveys/acs}} where about 9.8 million workers are employed as health care technicians and practitioners, including physicians, surgeons, and registered nurses, which accounts for 4\% of adults.

\begin{table}[h!]
  \begin{center}
    \caption{Performance of the medical professional inference.}
    \label{tab:keyword}
    \begin{tabular}{lr} 
    \hline
      \textbf{Metrics} & \textbf{Score}\\
      \hline
      F1 Score & 90.7\\
      Precision & 88.6\\
      Recall & 92.9\\
      \hline
    \end{tabular}
  \end{center}
\end{table}

Using these 148 keywords, we further classify the identified medical professionals into (1) doctors, (2) nurses, and (3) other medical professionals.

\subsection{Irrelevant Data Removal}
Our goal is to have a better understanding of how medical practitioners think about COVID-19 vaccines. However, similar to \citet{lyu2022social}, not every tweet we collect reveals users' opinions. They could simply be notifications related to vaccines. For example,\footnote{The example tweet has been paraphrased to protect user privacy.}

{\tt "Students should learn from home if the class does not have at least a 70\% vaccination rate where the green passport restrictions are applied."}

Therefore, we employ the XLNet model~\cite{yang2019xlnet} from \citet{lyu2022social}, where they fine-tuned the model with 4,900 manually labeled data. The data were acquired in the same way as we do. They labeled them as relevant or irrelevant based on whether or not, by manual inspections, a researcher thinks the tweet expresses pro-vaccine, vaccine-hesitant, anti-vaccine opinions. If the tweet does not express any of the opinions toward COVID-19 vaccines, it will be labeled irrelevant. The model fined-tuned with these data achieves good performance on the relevant/irrelevant classification task (weighted precision/recall/F1-score = 0.70/0.63/0.63).

Consequently, we apply this model to our data and remove all the irrelevant tweets. It is noteworthy that before we apply this model, we adopt a tweet preprocessing pipeline from \citet{baziotis2017datastories} which converts specific components such as handle names, URLs (Uniform Resource Locators), capitalized letters into special tokens. For example, if the original tweet is

{\tt "How the authoritarians among us will react to COVID vaccine mandates \url{https://t.co/...}"}

After the preprocessing pipeline, it becomes 

{\tt "how the authoritarians among us will react to $<allcaps>$covid$</allcaps>$ vaccine mandates $<url>$"}

\subsection{Sentiment Analysis}
In our study, we focus on investigating the sentiment toward the COVID-19 vaccines expressed by the healthcare workers. It is important to note that sentiment is not necessarily stance. According to \citet{lyu2022social}, the sentiment toward COVID-19 is positively correlated with the likelihood of holding pro-vaccine opinions. Therefore, we use sentiment as a proxy to estimate healthcare workers' opinions.

For each tweet, we infer its sentiment using VADER (Valence  Aware  Dictionary  for sEntiment Reasoning), a rule-based model for general sentiment analysis~\cite{hutto2014vader}. It calculates a normalized, weighted composite score to estimate the sentiment expressed by the tweet content. This score is between -1 (most extreme negative) and +1 (most extreme positive).  If the score is greater than or equal to 0.05, the tweet is labeled positive. If the score is lower than or equal to -0.05, it is labeled negative. Otherwise, the tweet is labeled neutral.

\subsection{Topic Modelling}
Topic modeling is a method for unsupervised classification of documents, which finds a similar group of items (topics). A document can be part of multiple topics. We intend to analyze the relationship between the sentiments and the topics. We apply Latent Dirichlet Allocation (LDA)~\cite{blei2003latent} to extract the topics from all relevant tweets. Similarly, we apply LDA to other four groups of tweets: (1) tweets posted by doctors with positive sentiments, (2) tweets posted by nurses with positive sentiments, (3) tweets posted by doctors with negative sentiments, and (4) tweets posted by nurses with negative sentiments. We choose the number of topics on the basis of both the coherence scores and manual inspections. The number of topics is set as five.

\begin{figure}
    \centering
    \includegraphics[width = \linewidth]{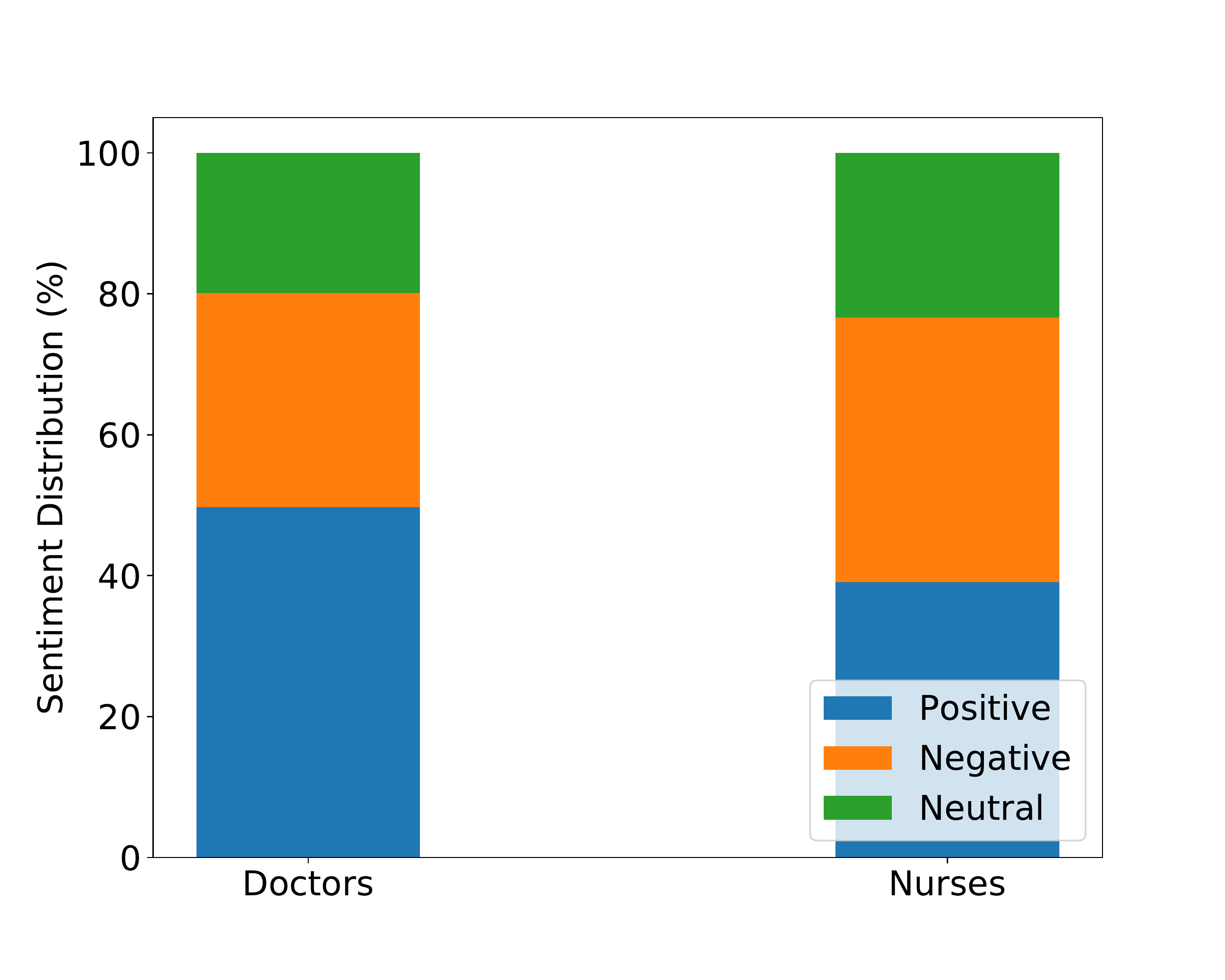}
    \caption{Sentiment distributions of the tweets posted by doctors and nurses.}
    \label{fig:sentiment_dist}
\end{figure}

\begin{figure}[t]
        \centering
        \includegraphics[width=\linewidth]{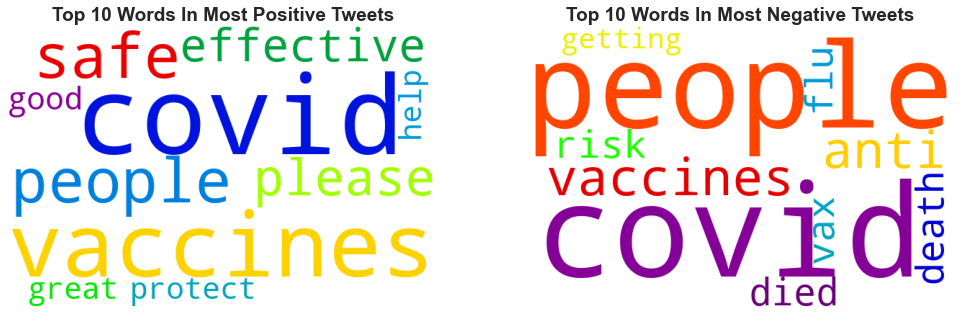} 
        \caption{Word clouds of the most positive and negative tweets posted by doctors.}
        \label{fig:tweet_doctor}
\end{figure}

\begin{figure}[t]
        \centering
        \includegraphics[width=\linewidth]{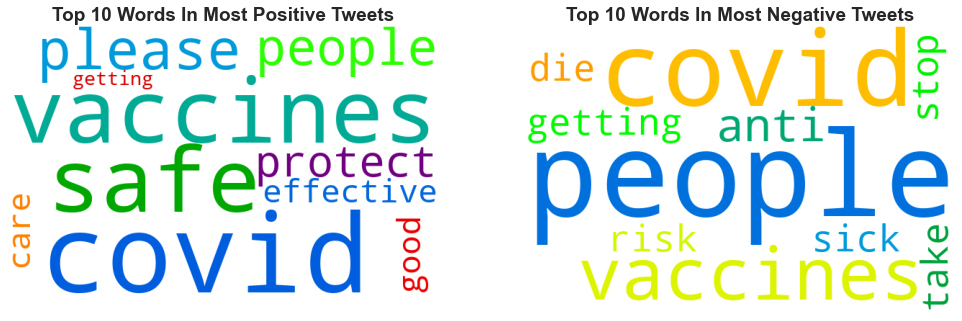} 
        \caption{Word cloud of the most positive and negative tweets posted by nurses.}
        \label{fig:tweet_nurse}
\end{figure}

\begin{figure*}[htbp]
\centering
\includegraphics[width=\linewidth]{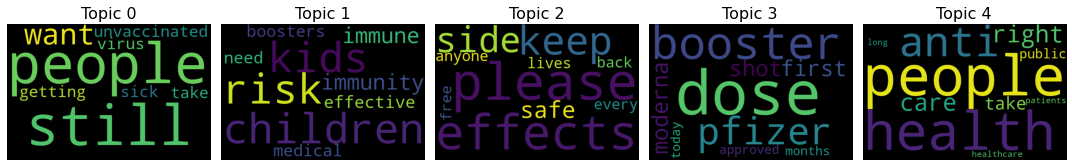} 
\caption{Word cloud of the top 10 keywords in each topic extracted by LDA on all tweets.}
\label{fig:all_topic}
\end{figure*}

\begin{table*}[htbp]
\small
    \centering
    \caption{Topics of positive/negative tweets posted by doctors or nurses.}
\begin{tabular}{llll}
\hline
 \textbf{Positive tweets of doctors}  & \textbf{Negative tweets of doctors }                     & \textbf{Positive tweets of nurses }& \textbf{Negative tweets of nurses} \\
\hline
Protect family and children & Ineffectiveness of vaccines on the new variants & Protect patients           & Risk                      \\
Booster                     & Right to not get vaccinated                     & Mask requirement          & Impact on children        \\
Immunity            & Long-term illness                               & Immunity        & Mandate                    \\
Safety/effectiveness        & Death/risk                                          & Safety/effectiveness      & Long-term illness         \\
Science              & Adverse affects of vaccines                     & Booster   & Government regulations   \\
\hline
\end{tabular}
    
    \label{tab:topics}
\end{table*}

\begin{figure*}[htbp]
    \centering
    \includegraphics[width =0.88 \linewidth]{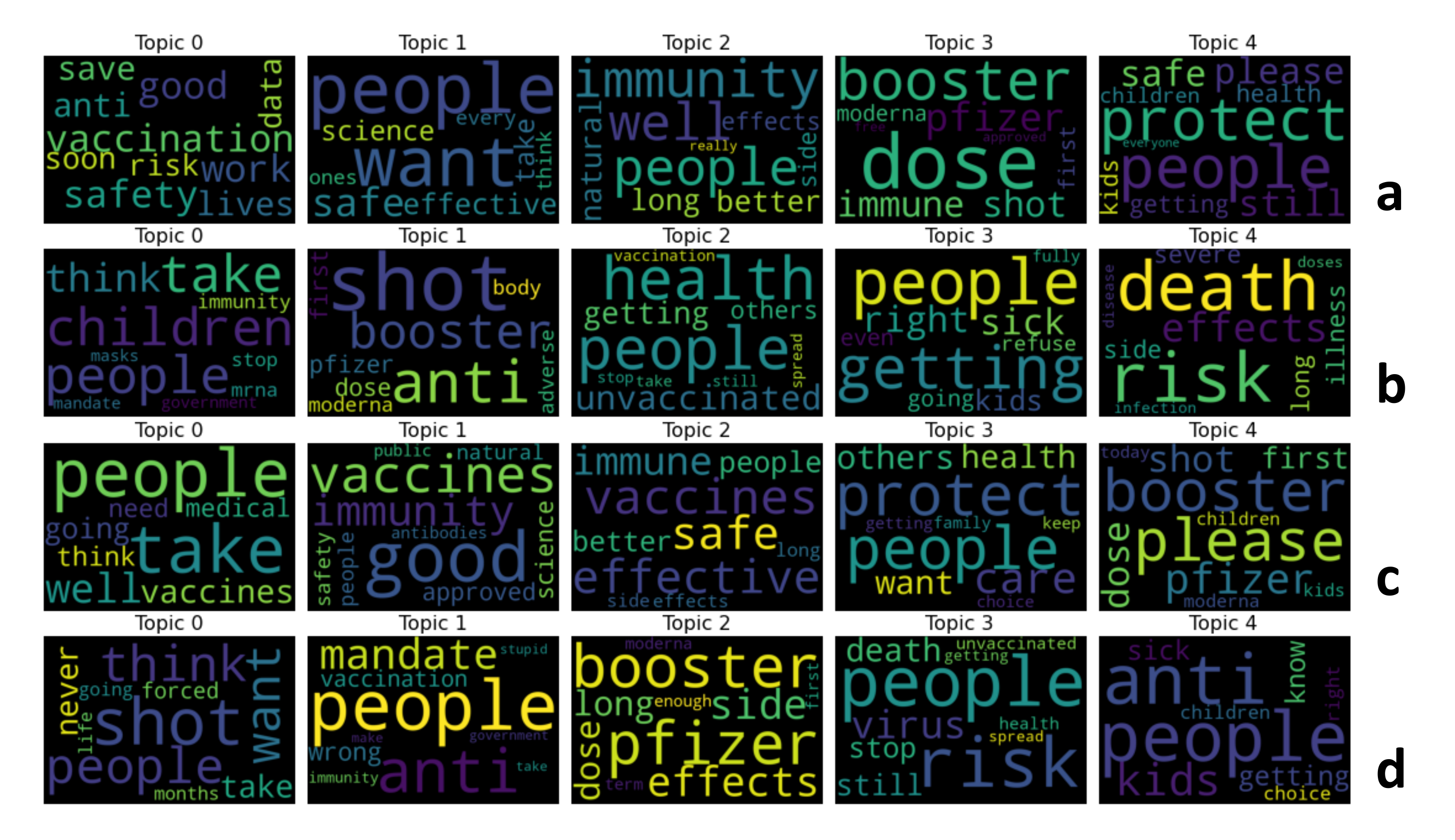}
    \caption{Top 10 keywords of each topic of the positive/negative tweets by doctors or nurses. Panels (a) and (b) show the keywords of the positive and negative tweets of doctors. Panels (c) and (d) show the keywords of the negative tweets of nurses.}
    \label{fig:topic_keywords}
\end{figure*}

\section{Results}
Figure~\ref{fig:sentiment_dist} shows the sentiment distributions of the tweets posted by doctors and nurses. Among all tweets that are posted by doctors, 49.7\% are positive, 30.4\% are negative, 19.9\% are neutral, while among all tweets that are posted by nurses, the proportions of positive, negative, and neutral classes are 39.1\%, 37.5\%, and 23.4\%, respectively. After performing the Chi-squared test, we find the sentiment distributions of the tweets posted by doctors and nurses are \textbf{significantly different} ($p<.05$). Overall, tweets posted by doctors show more positivity about the COVID-19 vaccines.

Figure~\ref{fig:tweet_doctor} shows the word clouds of the most positive and negative tweets posted by doctors. We find that doctors talk about the effectiveness of the vaccines and urge people to get vaccinated in the positive tweets. In contrast, the pessimistic tweets reveal the risk involved in getting vaccinated.

Figure~\ref{fig:tweet_nurse} shows the word clouds of the most positive and negative tweets posted by nurses. In the positive tweets, the nurses express their feelings about caring for people, requesting people to get vaccinated and stay safe from the virus. They are cautious about the risk and illnesses that might be the side effects of vaccines, which is depicted in the negative tweets.

 Figure~\ref{fig:all_topic} shows the top 10 keywords in each topic extracted by LDA on all tweets. There are a variety of topics including effectiveness (Topic 0), risk for children (Topic 1), side effects (Topic 2), different vaccines (Topic 3), and anti-vaccine sentiment (Topic 4).

Figure~\ref{fig:topic_keywords} shows the top 10 keywords of each topic of the positive/negative tweets posted by doctors or nurses. We assign a concise topic label to each of them to summarize the themes in Table~\ref{tab:topics}.

By analyzing the topics of the positive tweets, we observe that when talking positively about the COVID-19 vaccines, the focuses of doctors and nurses are similar. They both mention the safety and effectiveness, collective responsibility (i.e., the purpose of protecting others such as vulnerable children and patients). They discuss the immunity obtained by getting vaccinated. Further, doctors mention that about 95\% of COVID patients getting admitted to the hospitals are unvaccinated, urging the public to get both vaccinations and the booster. The nurses are more concerned with the safety of the children, point out that the requirement of masks will not be there if vaccinated, and argue that vaccines are for obtaining immunization from COVID-19. 

We also observe some common concerns of the doctors and nurses when they discuss COVID-19 in a negative way. The risk in getting vaccinated, mandates, and long-term illness are the most frequent topics among the negative tweets posted by doctors and nurses. In particular, doctors are skeptical about vaccine efficacy on newer virus variants. The nurses with negative sentiment are mostly concerned about the risks of the vaccine and the government's regulation. They are concerned about the well-being of their children after the vaccine is administered as they think there have not been enough tests.

\section{Discussion and Conclusion}
In this study, we use the keyword search to identify medical professionals on Twitter and collect their discussion about the COVID-19 vaccines. We use the XLNet model to retain the tweets that reveal their opinions. We measure the sentiment of the tweets posted by doctors and nurses using VADER to estimate the opinions and further conduct a thematic analysis.

We find that doctors are overall more positive toward the COVID-19 vaccines than nursers. This is \textit{consistent} with the findings of \citet{browne2021coronavirus} where they found nurses are more vaccine-hesitant. There are both positive and negative tweets posted by the doctors and nurses. One potential reason that doctors are more positive could be that they are trained to take the trade-off when the benefits outweigh the risks,\footnote{\url{https://www.cdc.gov/coronavirus/2019-ncov/vaccines/vaccine-benefits.html}} while the nurses do not and many even decide to quit their jobs to avoid vaccination.\footnote{\url{https://www.bloomberg.com/news/articles/2021-08-26/covid-vaccine-mandates-drive-some-nurses-to-leave-america-s-hospitals}} By further analyzing the topics in the tweets posted by doctors and nurses, we find that doctors and nurses focus on different aspects about the COVID-19 vaccines when discussing it negatively, although there are several common concerns such as the risks in getting vaccinated, mandates and long-term illness. This echos the findings of previous studies~\cite{adejumo2021perceptions, baniak2021covid}.

Some doctors feel their right to reject the vaccine is being violated, and they are being forced to take vaccines. They are skeptical about the efficacy of the vaccines on future variants and demand that more tests or trials to be done on the vaccines so that they do not lead to long-term illnesses. However, the nurses discuss more about their children's health and argue that vaccines should not be mandated on children. Nurses are also worried about the long-term effect of vaccines. This suggests that the various focuses might be the reason that the levels of vaccine hesitancy of the doctors and nurses are different. 

To increase vaccine uptake among healthcare workers, more customized strategies should be deployed. For example, more experiments results showing the effectiveness of vaccines or boosters on the new variants should be communicated with doctors. Experiment results of vaccination on children should be highlighted when communicating with nurses.

Our study has several limitations. First, we only focus on one social platform. Future work can analyze the social media posts across various platforms to obtain a more comprehensive understanding. Second, we use the keyword search method to detect medical professionals. This works well on Twitter, however, other methods may need to be developed on other platforms. Third, our work can be extended to examining the opinions of medical students.

\section{Acknowledgments}
This research was supported in part by the Goergen Institute for Data Science.

\bibliography{report}

\begin{thebibliography}{24}
\providecommand{\natexlab}[1]{#1}
\providecommand{\url}[1]{\texttt{#1}}
\providecommand{\urlprefix}{URL }
\expandafter\ifx\csname urlstyle\endcsname\relax
  \providecommand{\doi}[1]{doi:\discretionary{}{}{}#1}\else
  \providecommand{\doi}{doi:\discretionary{}{}{}\begingroup
  \urlstyle{rm}\Url}\fi

\bibitem[{Abd-Alrazaq et~al.(2020)Abd-Alrazaq, Alhuwail, Househ, Hamdi, Shah
  et~al.}]{abd2020top}
Abd-Alrazaq, A.; Alhuwail, D.; Househ, M.; Hamdi, M.; Shah, Z.; et~al. 2020.
\newblock Top concerns of tweeters during the COVID-19 pandemic: infoveillance
  study.
\newblock \emph{Journal of medical Internet research} 22(4): e19016.

\bibitem[{Adejumo et~al.(2021)Adejumo, Ogundele, Madubuko, Oluwafemi, Okoye,
  Okonkwo, Owolade, Junaid, Lawal, Enikuomehin et~al.}]{adejumo2021perceptions}
Adejumo, O.~A.; Ogundele, O.~A.; Madubuko, C.~R.; Oluwafemi, R.~O.; Okoye,
  O.~C.; Okonkwo, K.~C.; Owolade, S.~S.; Junaid, O.~A.; Lawal, O.~M.;
  Enikuomehin, A.~C.; et~al. 2021.
\newblock Perceptions of the COVID-19 vaccine and willingness to receive
  vaccination among health workers in Nigeria.
\newblock \emph{Osong Public Health and Research Perspectives} 12(4): 236.

\bibitem[{Amuzie et~al.(2021)Amuzie, Odini, Kalu, Izuka, Nwamoh, Emma-Ukaegbu,
  and Onyike}]{amuzie2021covid}
Amuzie, C.~I.; Odini, F.; Kalu, K.~U.; Izuka, M.; Nwamoh, U.; Emma-Ukaegbu, U.;
  and Onyike, G. 2021.
\newblock COVID-19 vaccine hesitancy among healthcare workers and its
  socio-demographic determinants in Abia State, Southeastern Nigeria: a
  cross-sectional study.
\newblock \emph{The Pan African Medical Journal} 40.

\bibitem[{Baniak et~al.(2021)Baniak, Luyster, Raible, McCray, and
  Strollo}]{baniak2021covid}
Baniak, L.~M.; Luyster, F.~S.; Raible, C.~A.; McCray, E.~E.; and Strollo, P.~J.
  2021.
\newblock COVID-19 vaccine hesitancy and uptake among nursing staff during an
  active vaccine rollout.
\newblock \emph{Vaccines} 9(8): 858.

\bibitem[{Baziotis, Pelekis, and Doulkeridis(2017)}]{baziotis2017datastories}
Baziotis, C.; Pelekis, N.; and Doulkeridis, C. 2017.
\newblock Datastories at semeval-2017 task 4: Deep lstm with attention for
  message-level and topic-based sentiment analysis.
\newblock In \emph{Proceedings of the 11th international workshop on semantic
  evaluation (SemEval-2017)}, 747--754.

\bibitem[{Blei, Ng, and Jordan(2003)}]{blei2003latent}
Blei, D.~M.; Ng, A.~Y.; and Jordan, M.~I. 2003.
\newblock Latent dirichlet allocation.
\newblock \emph{Journal of machine Learning research} 3(Jan): 993--1022.

\bibitem[{Browne et~al.(2021)Browne, Feemster, Shen, Green-McKenzie,
  Momplaisir, Faig, Offit, and Kuter}]{browne2021coronavirus}
Browne, S.~K.; Feemster, K.~A.; Shen, A.~K.; Green-McKenzie, J.; Momplaisir,
  F.~M.; Faig, W.; Offit, P.~A.; and Kuter, B.~J. 2021.
\newblock Coronavirus disease 2019 (COVID-19) vaccine hesitancy among
  physicians, physician assistants, nurse practitioners, and nurses in two
  academic hospitals in Philadelphia.
\newblock \emph{Infection Control \& Hospital Epidemiology} 1--9.

\bibitem[{Cucinotta and Vanelli(2020)}]{cucinotta2020declares}
Cucinotta, D.; and Vanelli, M. 2020.
\newblock WHO declares COVID-19 a pandemic.
\newblock \emph{Acta Bio Medica: Atenei Parmensis} 91(1): 157.

\bibitem[{De~Figueiredo et~al.(2020)De~Figueiredo, Simas, Karafillakis,
  Paterson, and Larson}]{de2020mapping}
De~Figueiredo, A.; Simas, C.; Karafillakis, E.; Paterson, P.; and Larson, H.~J.
  2020.
\newblock Mapping global trends in vaccine confidence and investigating
  barriers to vaccine uptake: a large-scale retrospective temporal modelling
  study.
\newblock \emph{The Lancet} 396(10255): 898--908.

\bibitem[{Gallotti et~al.(2020)Gallotti, Valle, Castaldo, Sacco, and
  De~Domenico}]{gallotti2020assessing}
Gallotti, R.; Valle, F.; Castaldo, N.; Sacco, P.; and De~Domenico, M. 2020.
\newblock Assessing the risks of ‘infodemics’ in response to COVID-19
  epidemics.
\newblock \emph{Nature human behaviour} 4(12): 1285--1293.

\bibitem[{Gao et~al.(2020)Gao, Zheng, Jia, Chen, Mao, Chen, Wang, Fu, and
  Dai}]{gao2020mental}
Gao, J.; Zheng, P.; Jia, Y.; Chen, H.; Mao, Y.; Chen, S.; Wang, Y.; Fu, H.; and
  Dai, J. 2020.
\newblock Mental health problems and social media exposure during COVID-19
  outbreak.
\newblock \emph{Plos one} 15(4): e0231924.

\bibitem[{Grumbach et~al.(2021)Grumbach, Judson, Desai, Jain, Lindan,
  Doernberg, and Holubar}]{grumbach2021association}
Grumbach, K.; Judson, T.; Desai, M.; Jain, V.; Lindan, C.; Doernberg, S.~B.;
  and Holubar, M. 2021.
\newblock Association of race/ethnicity with likeliness of COVID-19 vaccine
  uptake among health workers and the general population in the San Francisco
  Bay Area.
\newblock \emph{JAMA internal medicine} 181(7): 1008--1011.

\bibitem[{Hutto and Gilbert(2014)}]{hutto2014vader}
Hutto, C.; and Gilbert, E. 2014.
\newblock Vader: A parsimonious rule-based model for sentiment analysis of
  social media text.
\newblock In \emph{Proceedings of the international AAAI conference on web and
  social media}, volume~8, 216--225.

\bibitem[{Iliyasu et~al.(2022)Iliyasu, Garba, Gajida, Amole, Umar, Abdullahi,
  Kwaku, Salihu, and Aliyu}]{iliyasu2022should}
Iliyasu, Z.; Garba, M.~R.; Gajida, A.~U.; Amole, T.~G.; Umar, A.~A.; Abdullahi,
  H.~M.; Kwaku, A.~A.; Salihu, H.~M.; and Aliyu, M.~H. 2022.
\newblock ‘Why Should I Take the COVID-19 Vaccine after Recovering from the
  Disease?’A Mixed-methods Study of Correlates of COVID-19 Vaccine
  Acceptability among Health Workers in Northern Nigeria.
\newblock \emph{Pathogens and Global Health} 116(4): 254--262.

\bibitem[{Kara~Esen et~al.(2021)Kara~Esen, Can, Pirdal, Aydin, Ozdil, Balkan,
  Budak, Keskindemirci, Karaali, and Saltoglu}]{kara2021covid}
Kara~Esen, B.; Can, G.; Pirdal, B.~Z.; Aydin, S.~N.; Ozdil, A.; Balkan, I.~I.;
  Budak, B.; Keskindemirci, Y.; Karaali, R.; and Saltoglu, N. 2021.
\newblock COVID-19 vaccine hesitancy in healthcare personnel: A university
  hospital experience.
\newblock \emph{Vaccines} 9(11): 1343.

\bibitem[{Khubchandani et~al.(2021)Khubchandani, Sharma, Price, Wiblishauser,
  Sharma, and Webb}]{khubchandani2021covid}
Khubchandani, J.; Sharma, S.; Price, J.~H.; Wiblishauser, M.~J.; Sharma, M.;
  and Webb, F.~J. 2021.
\newblock COVID-19 vaccination hesitancy in the United States: a rapid national
  assessment.
\newblock \emph{Journal of Community Health} 46(2): 270--277.

\bibitem[{Lucia, Kelekar, and Afonso(2021)}]{lucia2021covid}
Lucia, V.~C.; Kelekar, A.; and Afonso, N.~M. 2021.
\newblock COVID-19 vaccine hesitancy among medical students.
\newblock \emph{Journal of Public Health} 43(3): 445--449.

\bibitem[{Lyu et~al.(2022)Lyu, Wang, Wu, Duong, Zhang, Dye, and
  Luo}]{lyu2022social}
Lyu, H.; Wang, J.; Wu, W.; Duong, V.; Zhang, X.; Dye, T.~D.; and Luo, J. 2022.
\newblock Social media study of public opinions on potential COVID-19 vaccines:
  informing dissent, disparities, and dissemination.
\newblock \emph{Intelligent medicine} 2(01): 1--12.

\bibitem[{Lyu, Zheng, and Luo(2022)}]{lyu2022misinformation}
Lyu, H.; Zheng, Z.; and Luo, J. 2022.
\newblock Misinformation versus Facts: Understanding the Influence of News
  regarding COVID-19 Vaccines on Vaccine Uptake.
\newblock \emph{Health Data Science} 2022.

\bibitem[{Nasir et~al.(2021)Nasir, Zaman, Majumder, Ahmed, Nazneen, Omar,
  Perveen, Farha, Zahan, Hossain et~al.}]{nasir2021perception}
Nasir, M.; Zaman, M.~A.; Majumder, T.~K.; Ahmed, F.; Nazneen, R.; Omar, E.;
  Perveen, R.~A.; Farha, N.; Zahan, T.; Hossain, M.~J.; et~al. 2021.
\newblock Perception, preventive practice, and attitude towards vaccine against
  COVID-19 among health care professionals in Bangladesh.
\newblock \emph{Infection and Drug Resistance} 14: 3531.

\bibitem[{Tsao et~al.(2021)Tsao, Chen, Tisseverasinghe, Yang, Li, and
  Butt}]{tsao2021social}
Tsao, S.-F.; Chen, H.; Tisseverasinghe, T.; Yang, Y.; Li, L.; and Butt, Z.~A.
  2021.
\newblock What social media told us in the time of COVID-19: a scoping review.
\newblock \emph{The Lancet Digital Health} 3(3): e175--e194.

\bibitem[{Wu, Lyu, and Luo(2021)}]{wu2021characterizing}
Wu, W.; Lyu, H.; and Luo, J. 2021.
\newblock Characterizing discourse about covid-19 vaccines: A reddit version of
  the pandemic story.
\newblock \emph{Health Data Science} 2021.

\bibitem[{Yang et~al.(2019)Yang, Dai, Yang, Carbonell, Salakhutdinov, and
  Le}]{yang2019xlnet}
Yang, Z.; Dai, Z.; Yang, Y.; Carbonell, J.; Salakhutdinov, R.~R.; and Le, Q.~V.
  2019.
\newblock Xlnet: Generalized autoregressive pretraining for language
  understanding.
\newblock \emph{Advances in neural information processing systems} 32.

\bibitem[{Zhang et~al.(2021)Zhang, Lyu, Liu, Zhang, Wang, Luo
  et~al.}]{zhang2021monitoring}
Zhang, Y.; Lyu, H.; Liu, Y.; Zhang, X.; Wang, Y.; Luo, J.; et~al. 2021.
\newblock Monitoring depression trends on twitter during the COVID-19 pandemic:
  observational study.
\newblock \emph{JMIR infodemiology} 1(1): e26769.

\end{thebibliography}

\end{document}